%% file: main.tex
\documentclass[conference]{IEEEtran}
\IEEEoverridecommandlockouts
\usepackage{cite}
\usepackage{amsmath,amssymb,amsfonts}
\usepackage{algorithmic}
\usepackage{graphicx}
\usepackage{textcomp}
\usepackage{xcolor}
\def\BibTeX{{\rm B\kern-.05em{\sc i\kern-.025em b}\kern-.08em
    T\kern-.1667em\lower.7ex\hbox{E}\kern-.125emX}}
    
\usepackage{flushend}
\usepackage{multirow}
\usepackage{hyperref}
\usepackage{glossaries}
\usepackage{tikz,pgfplots}

\begin{document}

\title{Verilay: A Verifiable Proof of Stake Chain Relay}

\author{\IEEEauthorblockN{Martin Westerkamp}
	\IEEEauthorblockA{\textit{Service-centric Networking} \\
		\textit{Technische Universit\"at Berlin}\\
		Berlin, Germany \\
		westerkamp@tu-berlin.de}
	\and
	\IEEEauthorblockN{Maximilian Diez}
	\IEEEauthorblockA{\textit{Hasso Plattner Institute} \\
		\textit{University of Potsdam}\\
		Potsdam, Germany \\
		maximilian.diez@alumni.uni-potsdam.de}
}

\maketitle

\begin{abstract}
Blockchain relay schemes enable cross-chain state proofs without requiring trusted intermediaries. This is achieved by applying the source blockchain's consensus validation protocol on the target blockchain.
Existing chain relays allow for the validation of blocks created using the Proof of Work (PoW) protocol.
Since PoW entails high energy consumption, limited throughput, and no guaranteed finality, Proof of Stake (PoS) blockchain protocols are increasingly popular for addressing these shortcomings.
We propose Verilay, the first chain relay scheme that enables validating PoS protocols that produce finalized blocks, for example, Ethereum 2.0, Cosmos, and Polkadot. 
The concept does not require changes to the source blockchain protocols or validator operations.
Signatures of block proposers are validated by a dedicated relay smart contract on the target blockchain.
In contrast to basic PoW chain relays, Verilay requires only a subset of block headers to be submitted in order to maintain full verifiability. This yields enhanced scalability.
We provide a prototypical implementation that facilitates the validation of Ethereum 2.0 beacon chain headers within the Ethereum Virtual Machine (EVM).
Our evaluation proves the applicability to Ethereum 1.0's mainnet and confirms that only a fraction of transaction costs are required compared to PoW chain relay updates.
\end{abstract}

\input{glossaries}
\section{Introduction}
The increasing utilization of blockchain networks has entailed novel implementations fostering high throughput and advanced scalability.
In addition, emerging consensus algorithms such as \gls{pos} mitigate the inherent immense power consumption of \gls{pow}-based networks.
However, with a growing number of blockchain networks,  the resulting fragmentation constitutes a pressing challenge.

One of the most prominent use cases driving blockchain utilization in recent years is Decentralized Finance (DeFi).
The deployment of a plethora of DeFi applications to Ethereum has resulted in congestion that calls for offloading to other blockchain networks that provide higher throughput or are less utilized.
Yet, as DeFi applications such as decentralized exchanges require sufficient liquidity, isolated deployments to competing networks are not sufficient.
Enabling cross-chain token transfers through wrapped tokens alleviates dependencies on a single blockchain network and increases overall throughput.
Liquidity can be provided where necessary and blockchain advances supporting novel use cases are supported.

Multiple approaches have been presented to enable cross-chain data exchange~\cite{buterin2016interoperability}.
First, the simplest approach is based on (semi) trusted intermediaries that mediate between blockchains to exchange information.
Various implementations of such notary schemes exist and usually imply economic rationality assumptions\cite{heiss2019oracles}.
Therefore, the reliability of transferred data is not reflected in the source blockchain's consensus, but it depends on external entities and their respective collateral.

The second approach targets use cases that do not require any knowledge about the source blockchain state on the target blockchain.
For instance, atomic swaps enable the exchange of assets stored on two distinct blockchains between two entities by applying hash-locks in combination with a commit-reveal scheme~\cite{Meyden2019AtomicSwaps}.
Both involved entities retrieve required information by observing transactions occurring on both blockchains.
Reproducing transactions or states across blockchain networks is not needed.
However, such schemes cannot cater for more elaborate use cases such as moving assets between blockchain networks and suffer from attacks exploiting time delays to gain economic benefits~\cite{Xu2021AtomicSwaps}.

In contrast, chain relays enable blockchain interoperability without requiring trusted intermediaries or economic assumptions and facilitate complex cross-chain use cases that utilize the state, transactions, or events of a remote blockchain~\cite{westerkamp2020zkRelay}.
The consensus rules of a primary blockchain are incorporated on a secondary blockchain, typically in a smart contract.
Resembling light client functionality, block headers are submitted and validated on the secondary blockchain, enabling the utilization of Merkle proofs to determine facts from the source blockchain.
As the validation is conducted within a smart contract, no trust in the submitting entity is required.
Hereby, chain relays enable a wide range of cross-chain applications such as wrapped token transfer, cross-chain contract execution, or smart contract synchronization. 

While multiple chain relays have been proposed, they target the validation of Nakamoto consensus protocols and their derivatives.
The majority of these implementations enables the validation of \gls{pow}-based algorithms~\cite{btcrelay,westerkamp2020zkRelay,frauenthaler2020ethrelay}.
Furthermore, Ga\v{z}i et. al proposed a sidechain protocol for \gls{pos} algorithms that are derived from Nakamoto consensus~\cite{Gazi2019}.
Yet, these proposals do not cater for emerging blockchain implementations such as Ethereum 2.0~\cite{Buterin2020POS}, Polkadot~\cite{Wood2016} or Cosmos~\cite{buchman2019gossip} that provide guaranteed finality.

We propose Verilay, a chain relay concept enabling interoperability between \gls{pos}-based blockchains that support guaranteed finality and any blockchain offering smart contract functionality.
In this paper, we first analyze different \gls{pos} algorithms and classify their properties in the context of chain relays.
We differentiate between Nakamoto-based consensus algorithms that follow the longest-chain rule and those that are based on \gls{pbft}.
To the best of our knowledge, we propose the first chain relay compatible with \gls{pbft}-based \gls{pos} blockchains.
Compared to previous chain relay proposals, Verilay provides enhanced scalability by skipping blocks of the source blockchain and replacing formerly relayed blocks while remaining full verifiability of the entire blockchain history.

To demonstrate the viability of our proposed system design, we provide and evaluate a prototypical implementation that establishes a bridge between Ethereum 2.0's beacon chain and any blockchain supporting the \gls{evm}.
As the \gls{evm} is supported by a multitude of blockchain implementations such as Hyperledger Burrow\footnote{\url{https://github.com/hyperledger/burrow}} or shards of Avalanche\footnote{\url{https://docs.avax.network/}}, Polkadot\footnote{\url{https://www.parity.io/blog/substrate-evm/}}, and Cosmos\footnote{\url{https://github.com/tharsis/evmos}}, Verilay has the potential to provide trustless interoperability between \gls{pos}-based blockchain networks.


\section{Proof of Stake Protocols}
In the context of blockchain technologies, consensus algorithms are utilized to agree on a shared state in a decentralized setting~\cite{nakamoto2008}.
Transactions are disseminated in the network and subsumed in blocks to guarantee a common order of execution.
One of the main challenges in anonymous and permissionless blockchain networks is leader election, that is the election of a successive block proposer.
The application of \gls{pos} constitutes a solution for selecting such block proposers~\cite{Badertscher2018OuroborosGenesis,Buterin2020POS}.

\textbf{Staking.} \acrfull{pos} protocols are a family of consensus protocols that require a security deposit of cryptocurrency for gaining the right to publish blocks.
Hereby, sybil attacks are mitigated, as staked deposits act as a limiting factor~\cite{Wood2016}.
Furthermore, honest behavior is incentivized, for instance by distributing rewards for correct participation in the protocol or by burning stake in case of disclosed misbehavior, also referred to as slashing~\cite{Buterin2017}.
Protocols that apply slashing when misbehavior is detected are called accountably safe in the context of this paper~\cite{Buterin2020POS}.
Staking, also referred to as bonding~\cite{Kwon2014}, describes the process of locking assets to participate in the consensus protocol~\cite{Buterin2017}.
Stakeholders can withdraw stake in a process called unbonding.
The unbonding period defines a duration participants have to wait before stake can be withdrawn.

Typically, the probability of becoming a block proposer is either proportional to bound stake~\cite{Kiayias2017Ouroboros,Wood2016} or a fixed amount of stake is required to participate~\cite{Buterin2020POS}.
\Gls{dpos} describes an extension that permits staking to vote for a delegate who participates in the consensus protocol on behalf of the nominator~\cite{Wood2016}.

\Gls{pos} consensus protocols suffer from the so-called nothing-at-stake problem that allows for long-range-attacks~\cite{Wood2016}.
\Gls{pow}-based protocols do not suffer from such attacks, as the creation of forks requires significant computational and thus economic investments.
The creation of an illegitimate fork becomes more infeasible with each appended block in the competing fork.
In comparison, creating blocks in \gls{pos} calls for the computation of signatures, a relatively cheap operation compared to \gls{pow} mining that may permit colluding validators to create forks of arbitrary length to force blockchain reorganizations.
Various approaches exist to tackle such an attack.
For instance, double signing can be detected as misbehavior and respective stake slashed~\cite{Buterin2020POS}.
Protocols that accept only recent blocks to prevent long-range-attacks are denoted weakly subjective~\cite{Deirmentzoglou2019Attack}.

\textbf{Time.} Many \gls{pos} protocols divide time into slots and epochs~\cite{Buterin2020POS,Stewart2020Grandpa,Kiayias2017Ouroboros}.
A slot is defined as a time period that contains at most one valid block.
An epoch refers to a number of consecutive slots with the same set of validators responsible for voting on blocks~\cite{Buterin2020POS}.
As such, each transition between two epochs can be considered a scheduled validator set change.
The validator change rate defines how much of the validator set may change at each scheduled validator set change.

\textbf{Finality.} Finality describes if a particular block is certain to be part of the consecutive blockchain and cannot be reverted.
Most consensus protocols either implement \mbox{(near-)}instant finality or probabilistic finality, also referred to as eventual consensus~\cite{Stewart2020Grandpa}.
\mbox{(Near-)}instant finality implies that a block will be final once the consensus process has finished for this specific block.
Probabilistic finality by contrast is initiated with the publication of a block and then approaches finality with the execution of the consensus process of the entire blockchain.
Since guaranteed finality allows the application of a wide variety of optimizations towards the chain relay, Verilay focuses on implementing these protocols.

\textbf{Forks.} Finality properties are closely related to the creation and selection of forks.
A fork is defined as the existence of multiple formally valid blocks in one or more consecutive slots and may result from network partitions or attacks such as long-range-attacks~\cite{Deirmentzoglou2019Attack}.
A protocol's fork choice rule has an integral effect on the conceptualization of a respective chain relay.
In this work, focus on \gls{pbft}-based algorithms that circumvent the appearance of forks for slots that contain finalized blocks.

\subsection{Nakamoto-inspired PoS}
In Nakamoto-inspired consensus protocols, the longest chain\footnote{In \gls{pow} protocols, the term "longest chain" is an abbreviation for the fork that subsumes the most aggregated work.} of blocks is the preferred fork~\cite{nakamoto2008}.
The likelihood of a block being permanently included in the agreed blockchain increases the more successive blocks are appended.
While first proposed in 2008 as part of the Bitcoin protocol, the concept has also been applied to \gls{pos} protocols~\cite{Kiayias2017Ouroboros}.
With Ouroboros Genesis, Badertscher et. al have proposed a protocol that utilizes the block density after two forks diverge to choose the accepted chain rather than the longest chain~\cite{Badertscher2018OuroborosGenesis}.
Hereby, long-range-attacks are mitigated and nodes that turn offline or join the network at a later stage are able to synchronize from the genesis block and do not require trusted checkpoints.
Thus, the trustless synchroniization of Nakamoto-based \gls{pos} provides an advantage compared to \gls{pbft}-instpired algorithms, as outlined in Section~\ref{sec:pbft}.

On-chain validation of a Nakamoto-inspired PoS protocol requires signature validation, fork handling, and the validation of validator sets.
As finality is only probabilistic for Nakamoto-inspired protocols, the chain relay has to be able to handle forks of arbitrary length.
In order to synchronize, the fork's length or density is taken into account.
The validation of validator set changes typically requires awareness of the entire blockchain, as leader election depends on preceding stake distribution and subsequent validator sets are not explicitly signed. 
\subsection{PBFT-inspired PoS}
\label{sec:pbft}
\acrfull{pbft} is a consensus algorithm that enables state machine replication surviving byzantine faults in a permissioned setting~\cite{Castro1999PBFT}.
Participants cast votes in multiple rounds to commit to a common state.
The protocol tolerates up to $\frac{1}{3}$ of byzantine participants.
\Gls{pbft}-inspired \gls{pos} algorithms such as Tendermint~\cite{Kwon2014} or Gasper~\cite{Buterin2020POS} apply a comparable voting mechanism to facilitate consensus in a permissionless setting.
Validators vote on their view of the current state by signing blocks.
Hereby, (near)-instant finality is provided since a block holding a sufficient amount of signatures from the responsible validator set cannot be reverted and is therefore deemed final~\cite{Stewart2020Grandpa}.
This feature represents an advantage over Nakamoto-based protocols, as no probabilistic assumptions are required.

However, this comes at the cost of limited verifiablity when participants temporarily leave the network or join at a later stage, as fork choice depends on validators' votes.
Accountably safe and weakly subjective protocols prevent long-range attacks by slashing their stake in case double signing of multiple forks is disclosed.
Yet, stake can be withdrawn after the unbonding period elapsed, permitting adversaries to collude on double signing at practically no cost.
In a weakly subjective PoS protocol, signatures by validators are only trusted as long as validators are accountable~\cite{Deirmentzoglou2019Attack}.
The time period during which validators are accountable is called trusting period and must be shorter than the unbonding period~\cite{Braithwaite2020TendermintLightClient}.
In a weakly subjective consensus protocol, participants who have not received and validated any block for longer than the trusting period need to obtain a trusted state from a trusted source.
Based on this state, they can continue validating succeeding blocks.
Consequently, clients cannot synchronize starting from the genesis block after the first trusting period has elapsed if the respective blockchain applies a weakly subjective consensus protocol.

Accountable safety depicts a challenge for chain relays, as the property can only be maintained if valid block headers are submitted regularly.
For instance, a validator could double sign a fork from the main chain and submit it to the chain relay.
In case the trusting period has elapsed, the adversary could not be held accountable.
If the trusting period has not elapsed, accountablility can only be ensured in case observers capture potential misbehavior and report it to the source chain.
\section{Proof of Stake Chain Relay}

\subsection{Design Goals}
\label{sec:design_goals}
In order to guide the chain relay design, we introduce six design goals that are derived from related cross-chain literature and adapted to suit a \gls{pos} chain relay~\cite{Zamyatin2019,westerkamp2020zkRelay,Gazi2019,frauenthaler2020ethrelay,buterin2016interoperability,btcrelay}.
These design goals will be utilized to evaluate the proposed concept in Section~\ref{sec:evaluation}.
\begin{itemize}
    \item \textbf{Forkless.} The chain relay does not require any soft, hard, or velvet fork on the source blockchain or explicit collaboration of a subset of its validators~\cite{Zamyatin2019,westerkamp2020zkRelay,Gazi2019}.
    \item \textbf{Trustless.} There are no trusted intermediaries.
    Trust in stored block headers should be derived solely from the source blockchain's consensus validation.
    Any trusted state required for bootstrapping the chain relay can be verified independently by its users~\cite{westerkamp2020zkRelay,frauenthaler2020ethrelay,buterin2016interoperability}.
    \item \textbf{Autonomous.} As long as the source and target blockchains are active, the chain relay can be operated independently of any single entity~\cite{btcrelay}.
    \item \textbf{Robust.} The chain relay retains its state and can still be updated, even after extended periods of inactivity due to situations in which no updates are performed, for example if the target blockchain is congested~\cite{Deirmentzoglou2019Attack}.
    \item \textbf{Corresponding.} Only those block headers that are valid according to the source blockchain's consensus rules are processed by the chain relay~\cite{westerkamp2020zkRelay,frauenthaler2020ethrelay}.
    The replicated data should allow for inclusion proofs of transactions, states and events that occurred on the source blockchain.
    \item \textbf{Lightweight.} Executing the chain relay client should be efficient in terms of computation and memory usage.
    Corresponding update transactions should remain at least within the execution limits of the target ledger~\cite{westerkamp2020zkRelay,frauenthaler2020ethrelay}.
\end{itemize}
\subsection{Proof of Stake Relay}
In this section, we introduce Verilay, a chain relay enabling the validation of \gls{pbft}-based \gls{pos} blockchains.
We first introduce the generally applicable chain relay concept and exemplify its realization given the specifics of Ethereum 2.0's Gasper protocol~\cite{Buterin2020POS} in the succeeding section.
The validation process is applicable to any blockchain supporting (quasi) Turing-complete user-defined computations such as the \gls{evm}~\cite{Wood2014}.
Therefore, it is independent of the target blockchain's consensus algorithm and suitable to a wide range of blockchain networks.
Subsequently, we depict the required steps for relaying \gls{pos} block headers and outline the relay states and preconditions.

\textbf{Initialization.}
Any client attempting to synchronize with a \gls{pos}-based blockchain that is weakly subjective has to be bootstrapped with a trusted initial state.
The initial state consists of a trusted block and the epoch's validator set.
As this condition also applies to \gls{pos} chain relays, users must verify the initial state before utilizing the contract.
This process is similar to the initial verification of an unknown contract's logic to prevent unexpected behavior during its execution.
While the verification process is left to the skilled user, increasing utilization and security audits conducted by trusted entities may render the contract trustworthy to any user.

\textbf{Relay update.}
After the relay contract's deployment, any entity connected to source and target blockchain is enabled to update the relay state.
As the validation of submitted block headers is conducted in the relay contract, no trust in the relayer is required.
To apply an update, the Verilay client retrieves the destination block header, current validation set, and next validation set, prepares a transaction and submits it to the relay contract on the target blockchain.

\textbf{Finalized blocks.} 
The proposed chain relay only accepts blocks that are finalized according to the source blockchain's consensus rules.
Since finalized blocks are guaranteed to be part of the main chain, forks cannot occur and hence do not have to be handled by the relay contract.
Depending on the source blockchain's consensus algorithm, the finalization is accompanied by a delay, as validator's votes are collected in successive blocks~\cite{Buterin2020POS}. 

\textbf{Validator set.}
The size of the validator set varies greatly depending on the implementation.
While Cosmos envisions validator set sizes in the order of hundreds and Polkadot plans to scale up to one thousand, Ethereum 2.0 does not define an upper bound. 
The chain relay design must cater for these configurations.
As the validation of signatures requires both computing resources and available public keys, the costs for updating the relay increase with the number of validators.
Hence, it is infeasible for a relay contract that is executed in a constrained execution environment to process all signatures of the Ethereum 2.0 validator set.

To facilitate light client validation, Ethereum's first beacon chain upgrade called \emph{Altair} implemented so-called \emph{sync committees}~\cite{Ethereum2Specs2021}.
A sync committee is a periodically sampled subset of the full validator set and consist of 512 validators who generate a supplementary signature of each block that is used for synchronization by resource constrained devices.
Furthermore, sync committees are stable during a sync committee period, which is two orders of magnitude longer than a validator set epoch, and thus requires less frequent updates.
Successive sync committees are sampled one period in advance.
Since every block references both the currently valid and the the following sync committee, the transition between sync committees can be validated by light clients as long at least one block is submitted within each sync committee period.

Polkadot and Cosmos do not implement the concept of sync committees, because the election of a validator set natively selects a subset of stakers for validation~\cite{Wood2016,Cosmos2019Whitepaper}.
Thus, the set sizes are limited and signature validation is tolerable to light clients.
In the following, we presume the utilization of the full validator set where its size is limited and a sub-sample such as a sync committee otherwise.

\begin{table}[htbp]
\centering
\caption{Required data for a chain relay update in generic case and specific to Ethereum 2.0}
\label{tab:relay-update}
\renewcommand{\arraystretch}{1.3}
\begin{tabular}{p{0.5cm}|p{2cm}p{5cm}}
    &\textbf{Generic} & \textbf{Ethereum 2.0} \\ \hline 
    \multicolumn{1}{c|}{\multirow{5}{*}{\rotatebox[origin=c]{90}{Every update}}}&Finalized block header & Finalized block header root $b_{target}$ \\
    &\multirow{3}{2cm}{Finality proof or signature} & Latest block header root $b_{latest}$\\
    && Merkle proof that $b_{target}$ is final\\
    && Multi-signature by $v_{trusted}$/$v_{trustedNext}$ \\ \hline
    \multirow{4}{*}{\rotatebox[origin=c]{90}{\parbox{1.5cm}{\centering Validator set update}}} & Next validator set & Next sync committee $v_{next}$ \\
    & Next validator set proof or signature & Merkle proof that $v_{next}$ is the next validator set according to $b_{target}$\\
\end{tabular}
\end{table}

\textbf{Block validation.}
To validate a block, the minimum number of signatures must be checked, according to the consensus protocol.
In case of Ethereum 2.0, it is verified whether the sync committee's aggregated signature was generated by a threshold of its members.
Other implementations that do not apply signature aggregation require a sufficient amount of signatures to be available.
Therefore, transactions updating the relay contract must submit the finalized block header that is intended to be stored and a signature issued by the respective validator set, as depicted in Table~\ref{tab:relay-update}. 

If the signature validation succeeds, the chain relay substitutes the currently stored block with the newly validated block.
Hereby, storage costs are reduced, as allocating new storage is usually the most expensive operation in blockchains' execution environments ~\cite{Wood2014}.
Unlike Bitcoin or Ethereum 1.0, blocks in Ethereum 2.0 contain Merkle roots that reference the complete history.
Thereby, historical states are retrievable based on a single subsequent block.

\begin{figure*}[!t]
    \centering
    \includegraphics[width=0.98\linewidth]{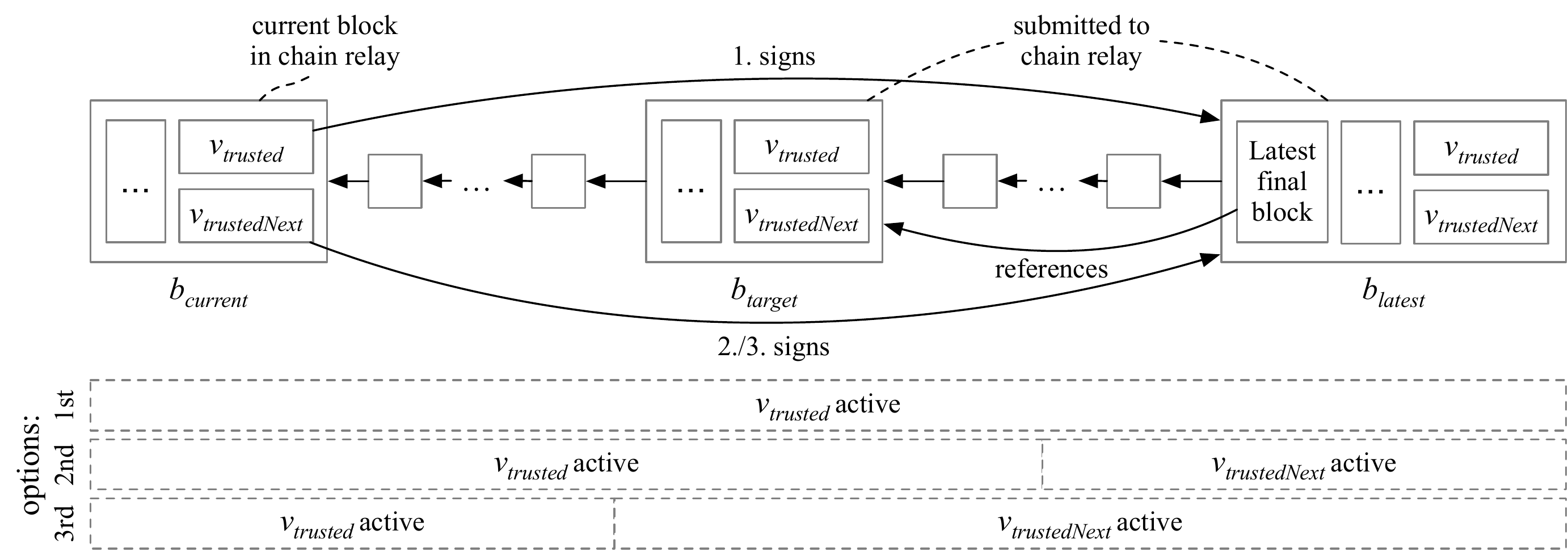}
    \caption{The chain relay always maintains one block header that is currently active. Two block headers must be submitted to update the relay contract, the target block and the latest block that finalizes it. The three involved block headers may be part of different sync committee periods, determining which sync committee signs the latest block and if the sync committee transitions.}
    \label{fig:trust_period}
\end{figure*}

In addition, if the new block is part of a new epoch, the transition between both validator sets is verified and the old set is replaced by the new one.
New validator sets are either signed in the previous epoch (Ethereum 2.0) or share significant intersections that can be used to verify transitions.
For instance, in the case of Cosmos, its security model assumes a minimum of $\frac{1}{3}$ of honest validators.
Submitted block headers must be within a certain bound.
According to the light client synchronization protocol of Cosmos~\cite{Braithwaite2020TendermintLightClient}, a block successive block can be validated by the chain relay as long as the previously stored block was signed by an at least $\frac{1}{3}$ of equivalent validators. 
In case of Ethereum 2.0, at least one block out of each epoch is required to verify the transition between validator sets.
Every chain relay update transaction that includes a block header which was signed by a new validator set must also include the subsequent validator set and a proof or signature attesting the transition.
Table~\ref{tab:relay-update} presents the required parameters for chain relay updates within a single epoch and across epochs.

\textbf{Accountable safety.}
Provided that the source blockchain supports accountable safety, the chain relay should retain this property.
As accountable safety can only be ensured within the trusting period of the signing validator set, updates must be submitted to the chain relay before the trusting period elapses.
Otherwise, the chain relay expires and must be redeployed including an initial trusted validator set.
This is due to potentially unbonded stake that permits colluding validators to sign blocks of multiple forks without being held accountable.
To retain accountable safety, observers must additionally detect misbehavior and submit respective proofs to the source blockchain that applies slashing.

If the source blockchain protocol does not include slashing, the chain relay does not need to adhere to such boundaries.
Some protocols such as Etherum 2.0 apply accountable safety to full validation, but not for sub-samples such as sync committees, as misbehavior in such a subset would provide only limited guarantees of accountability.
Thus, the use of sync committees for chain relay updates does not supply accountable safety.

\subsection{Ethereum 2.0 Relay}
The consensus algorithm Gasper that is utilized by Ethereum 2.0 implements two distinct concepts for block production and finalization respectively, namely LMD GHOST and \gls{pbft} inspired Casper FFG~\cite{Buterin2020POS}.
Casper FFG finalizes checkpoints by collecting votes in two rounds that first justify a checkpoint before finalizing it.
This process is derived from the commit phases of \gls{pbft} and adapted to a permissionless setting.
In Ethereum 2.0, a checkpoint is defined as the first block of an epoch and implicitly finalizes all preceding blocks.
Since votes are collected in succeeding blocks, checkpoints are finalized as soon as a sufficient amount of votes has been collected.
As Verilay aims for providing finalized blocks, two blocks must be submitted for updating its state, the finalized block $v_{target}$ which is to be stored by the relay contract and the latest block $v_{latest}$ that  declares $v_{target}$ to be finalized.
Every block contains a reference to the currently active sync committee $v_{trusted}$ and the subsequent sync committee $v_{trustedNext}$.
The relay contract verifies $b_{latest}$ by validating the signatures according to either $v_{trusted}$ or $v_{trustedNext}$ that are comprised in the currently stored block $b_{current}$.
Depending on the sync committee periods the blocks fall into, three different scenarios need to be considered, as illustrated in Figure~\ref{fig:trust_period}.

In the first case, all three blocks $b_{current}$, $b_{target}$ and $b_{latest}$ are part of the same sync committee period.
Therefore, the public keys of sync committee $v_{trusted}$ included in $b_{current}$ are used to validate the block signature for $b_{latest}$.
As $b_{current}$ has been successfully verified before, the referenced sync committee is trusted as well.
Since no sync committee update is needed, only the block headers $b_{target}$ and $b_{latest}$, a Merkle proof that $b_{target}$ is final according to $b_{latest}$ and a multi-signature by $v_{trusted}$ are required to update the chain relay, as depicted in Table~\ref{tab:relay-update}.

In the second case, $b_{current}$ and $b_{target}$ are contained in sync committee period $v_{trusted}$, while $b_{latest}$ is part of the successive sync committee period $v_{trustedNext}$.
Thus, $v_{trustedNext}$ referenced in $b_{current}$ must be used to verify the multi signature attesting the validity of $b_{latest}$.
Because $b_{current}$ and $b_{target}$ reference equivalent sync committees, there is no transition to a newly active sync committee.
Even though $b_{latest}$ includes a new sync committee reference, it is not applied, as the block has not necessarily been finalized at the time of submission.
As a result, compared to case one, the same data must be passed to apply an update, but a different validator set is used for signature validation.

Lastly, $b_{current}$ may be part of $v_{trusted}$, while $b_{target}$ and $b_{latest}$ are included within $v_{trustedNext}$.
Here, signature validation is performed analogously to the one in case two.
However, a transition to a new validator set occurs, transitioning former $v_{trustedNext}$ to $v_{trusted}$ and accepting a new set for $v_{trustedNext}$.
Thus, in addition to the the data that must be submitted for every relay update, the succeeding sync committee $v_{next}$ and a Merkle proof that it is part of $b_{target}$ must be passed, as detailed in Table~\ref{tab:relay-update}.
%
\section{Implementation}
\label{sec:implementation}
We provide a prototypical implementation of Verilay that enables the validation of Ethereum 2.0 beacon chain headers on any \gls{evm}-compatible blockchain.
Verilay is seperated into the relay contract that is hosted on the target blockchain and an off-chain client.
While the relay contract is responsible for validation, the client retrieves block headers from the beacon chain and prepares update transactions including respective proofs.
No authorization is required to update the relay state, as submitted block headers are validated by the relay contract.
Any user motivated for updating the relay contract requires access to a synchronized node of the source and target blockchain.
Verilay is available on GitHub\footnote{\url{https://github.com/MaximilianDiez/PoSChainRelay}} under an open-source license.

Each operation of the \gls{evm} has a cost assigned to it that is measured in \emph{gas} and payed by the transaction sender using the host blockchain's native currency~\cite{Wood2014}.
To reduce costs and enable lightweight updates, the Verilay contract requires only fractions of a block to be submitted while maintaining full verifiablity.
Referenced parts are validated using Merkle proofs that must be passed with every update.
In contrast to Ethereum 1.0 that uses a modified Merkle Patricia Tree to represent block data, Ethereum 2.0 implements a concept called \gls{ssz} for serializing data containers that are represented in a binary Merkle tree~\cite{Ethereum2Specs2021}.
As \gls{ssz} utilizes SHA256, which is available as cost-effective pre-compiled contract on Ethereum 1.0, for computing the Merkle tree, respective proofs can be verified efficiently on-chain.

\begin{figure}
    \centering
    \includegraphics[width=0.85\linewidth]{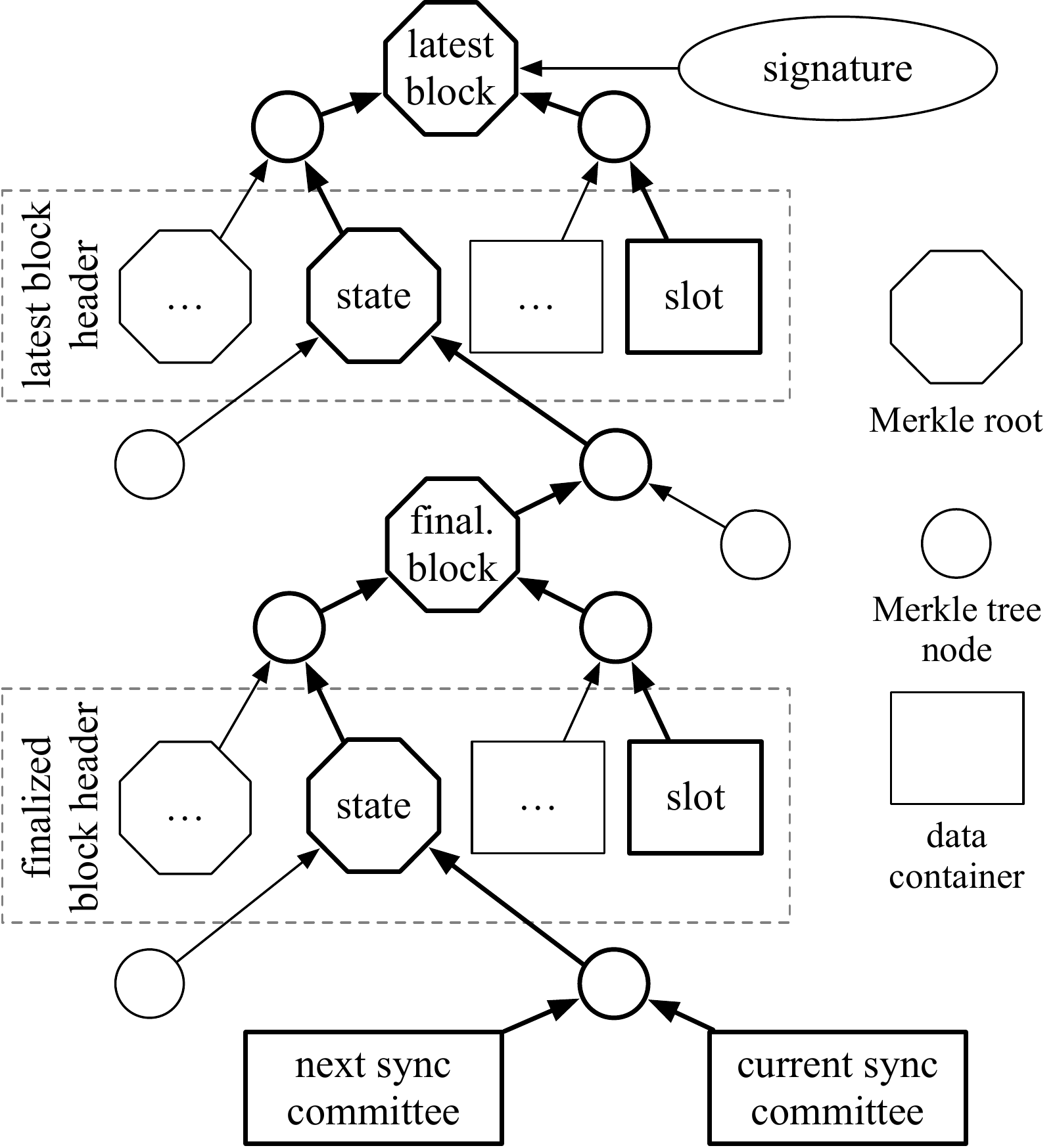}
    \caption{Simplified representation of required parameters in SSZ structure. Bold borders and arrows represent the Merkle paths of involved attributes.}
    \label{fig:ssz}
\end{figure}

The Verilay client prepares the parameters according to the requirements depicted in Table~\ref{tab:relay-update}.
Both the latest and the finalized block are represented by their Merkle root subsuming their referenced content.
The finalized block is referenced by the latest submitted block as part of the current state, as illustrated in Figure~\ref{fig:ssz}.
Since the sync committee used for validation is selected based on the block's slot, the slot is passed along with a respective Merkle proof of inclusion for both latest block and finalized block.

Verilay provides two options for accessing the sync committee's public keys during validation.
Option one requires passing the sync committee's public keys with any update transaction that results in the transition to a new sync committee.
Thereafter, the relay contract validates the respective Merkle proof and replaces the stored keys of the current sync committee with the new ones.
In Ethereum 2.0, the sync committee consists of 512 members.
Each member holds a public keys that $48\, byte$ long, resulting in a total size of $512 * 48\,byte = 24,576\,byte$.
As the update of a $32\,byte$ storage slot calls for $5,000\,gas$, the lower bound for updating the sync committee is $\frac{24,576\,byte}{32\,byte}*5,000\,gas=3,840,000\,gas$, excluding validation, data processing and memory operations.
The stored keys are retrieved from storage and used for each update's validation within in the same sync committee period.

As storage operations require comparatively large amounts of gas, we provide a second option that does not store the sync committee in the contract, but demands all public keys to be resubmitted with each update.
The sync committee period only determines which public key to pass and validate, but does not affect the transaction's number of attributes.
Resubmission of data to reduce storage costs has been proposed in literature before~\cite{Stelios2020LightClient,frauenthaler2020ethrelay}, but involved the calculation of a hash or Merkle root on-chain.
As the sync committee is part of the \gls{ssz} Merkle tree, this step can be omitted by the Verilay contract.
In Section~\ref{sec:evaluation}, we show that option two is beneficial even when updates are submitted regularly.

Multiple steps are required to verify the signature of the submitted latest block.
First, the relay contract checks whether a sufficient amount of sync committee members have participated.
Ethereum 2.0 enables signature aggregation by applying \gls{bls} multi-signatures based on the \gls{bls}12-381 elliptic curve~\cite{Boneh2001BLS,Ethereum2Specs2021}.
Operations on the \gls{bls}12-381 curve are conducted on 384-bit words.
Since the \gls{evm} operates on 256-bit words, the operations required to validate \gls{bls} signatures can not be executed efficiently by the virtual machine.
To provide an efficient alternative, a proposal\footnote{\url{https://eips.ethereum.org/EIPS/eip-2537}} has been published to add a pre-compiled contract to Ethereum 1.0 that executes \gls{bls}12-381 curve operations outside the \gls{evm}.
The pre-compiled contract will allow the efficient implementation of \gls{bls} signature validation within a smart contract.

As neither the proposed pre-compiled contract nor an on-chain signature validation library is available at the time of writing, we extended the Go Ethereum (geth) client with a customized pre-compiled contract that conducts validations using the Herumi BLS library\footnote{\url{https://github.com/herumi/bls}}.
To validate a submitted block header, the Verilay contract first serializes the sync committee's public keys and the aggregated signature, before calling the pre-compiled contract.
If the signature is valid, all Merkle proofs are checked to ensure the correct finalized block is referenced and correct slots have been declared.
Finally, the currently stored block header is replaced with the submitted finalized block header. 
%
\section{Evaluation}
\label{sec:evaluation}
In this section, we evaluate Verilay according to its quantitative properties such as on-chain execution costs and design design goals as defined in Section~\ref{sec:design_goals}. 
\subsection{Quantitative Analysis}
The chain relay's applicability depends in particular on the complexity of the associated on-chain computations.
Complex computations and high storage volumes increase transaction costs and may render transactions impossible to be executed in case they exceed the limits of a block on the target blockchain.
Therefore, we analyze transaction costs for deploying the relay contract and applying regular updates.
Since our chain relay prototype operates on the \gls{evm}, we measure transaction costs in gas, as defined by the Ethereum protocol~\cite{Wood2014}.

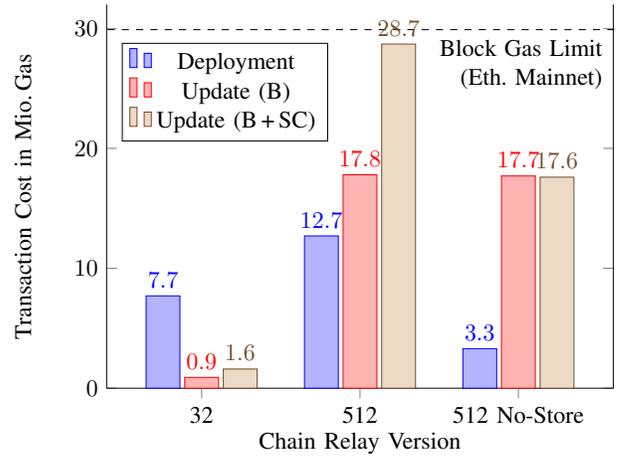
\begin{figure}[t!]
    \centering
    \scalebox{0.9}{
        \begin{tikzpicture}
        \draw[dashed] (7.5,5.3) -- (0,5.3);
        \node[anchor=south west, text width=2.5cm, align=right] () at (4.7,4.3) {Block Gas Limit (Eth. Mainnet)};
        \begin{axis}[
                ybar,
                bar width=.5cm,
                width=0.5\textwidth,
                height=.4\textwidth,
                symbolic x coords={32,512,512 No-Store},
                xtick={32,512,512 No-Store},
                enlarge x limits=0.3,
                axis x line*=bottom,
                nodes near coords,
                ymin=0,ymax=32,
                ylabel={Transaction Cost in Mio.\,Gas},
                xlabel={Chain Relay Version},
                legend style={at={(0.03,0.9)},anchor=north west}
            ]
            \addplot table[x=config,y=deployment]{\gasmeasurements};
            \addplot table[x=config,y=updateblock]{\gasmeasurements};
            \addplot table[x=config,y=updatesc]{\gasmeasurements};
            \legend{Deployment, Update (B), Update (B\,+\,SC)}
        \end{axis}
        \end{tikzpicture}
    }
    \caption{Measurements of the gas costs of deploying and updating different chain relay configurations. 32, 512 and 512 No-Store refer to chain relay versions for different sync committee sizes and whether the No-Store optimization was implemented. The current block gas limit on the Ethereum mainnet is indicated. B: Block. SC: Sync Committee.}
    \label{fig:gas-measurements}
\end{figure}

While the Ethereum 2.0 specifications define a sync committee size of 512, we also measure the costs of smaller sizes that result in reduced costs.
Even though our evaluation proves the applicability of Verilay for standard configurations of Ethereum 1.0 and 2.0, smaller sets may be applicable depending on the case's security requirements.
Figure~\ref{fig:gas-measurements} depicts the gas costs of three different configurations, using a reduced sync committee size of 32, applying the full set with 512 members while storing the set within the contract and resubmitting the set with every update, as described in Section~\ref{sec:implementation}.

We observe that the one-time deployment of the contract remains within the gas limits of the Ethereum mainnet.
The most gas is accounted for allocating storage for the sync committee.
Since the option to resubmit the set with every update does not require this step, the respective contract deployment is significantly cheaper, as illustrated in Figure~\ref{fig:gas-measurements}.
Furthermore, all operations on the smaller set are cheaper, as less storage operations and signature aggregations are required.
The most expensive transaction is an update that includes a sync committee transition, as the respective public keys must be passed and stored.
As expected, transitions of the sync committee are significantly cheaper when resubmitting the sync committee's public keys with every update, as no storage write operations are required.
Interestingly, update transactions that do not include such transitions also call for less gas compared to stored public keys.
Here, our results deviate from the evaluation of a similar approach taken by ETH Relay~\cite{frauenthaler2020ethrelay}.
We attribute this deviation to the Istanbul Hard Fork that included relevant changes to gas metering.
First, \gls{eip}-1884\footnote{\url{https://eips.ethereum.org/EIPS/eip-1884}} increased the costs for reading a storage value from 200 gas to 600 gas.
Second, \gls{eip}-2018\footnote{\url{https://eips.ethereum.org/EIPS/eip-2028}} reduced the price for passing call data from 68 gas to 16 gas.
We conclude that the described pattern is favorable when utilizing externally submitted data, even when regular resubmission is required.

\pgfplotstableread[row sep=\\,col sep=&]{
    config          & deployment    & updateblock    & updatesc \\
    32              & 7.7       & 0.9         & 1.6  \\
    512             & 12.7      & 17.8       & 28.7 \\
    512 No-Store    & 3.3       & 17.7       & 17.6 \\
    }\gasmeasurements

\subsection{Design Goals}
\label{sec:quantitative}
In the following, we analyze Verilay's compliance with the design goals introduced in Section~\ref{sec:design_goals} based on the presented prototype.
\begin{itemize}
    \item \textbf{Forkless.} Verilay utilizes sync committees that are part of the Ethereum 2.0 protocol and does not require collaboration of validators that is explicit to the chain relay.
    \item \textbf{Trustless.} After bootstrapping the relay contract, trust is solely derived from the source blockchain's sync committee's signature.
    As the validation is conducted on-chain, no trust in entities executing updates is needed.
    \item \textbf{Autonomous.} Anyone having access to the source and target ledger is enabled to submit updates by executing the Verilay client.
    \item \textbf{Robust.} While Ethereum 2.0 applies a weakly subjective consensus algorithm, sync committees do not hold this property.
    Hence, its state can be updated after arbitrarily long periods of inactivity.
    The relay may only stall in case no valid sync committee signature is present for an entire sync committee period.
    \item \textbf{Corresponding.} Since block headers are validated by sync committee members and a majority of at least $\frac{2}{3}$ is assumed to be honest, relayed block headers are also assumed to be valid.
    Merkle branches can be used to prove the inclusion of transactions, states and events.
    \item \textbf{Lightweight.} Verilay is lightweight as only a single block must be submitted for each sync committee period.
    Since such period subsumes 256 epochs of 32 slots, only a single update is required every 8,192 blocks, involving 17.7 million gas.
    In comparison, a single block update using BTC Relay calls for 194,000 gas\footnote{Exemplary transaction ID of a Bitcoin header submission in BTC relay: \href{https://etherscan.io/tx/0xe21099d8fd1252281389fc888f23f98e60db22ecb5c149ad6fda6dccdf110b50}{0xe21099d8fd1252281389fc888f23f98e60db22ecb5c149ad6fda6dccdf110b50}}, resulting in 1,589,248,000 gas for an equivalent number of blocks.
\end{itemize}


\section{Related Work}
%
Multiple chain relay schemes have been proposed performing on-chain validation.
BTC Relay was the first chain relay to be deployed in a live setting and enables the submission of Bitcoin headers to Ethereum to prove the inclusion of transactions~\cite{btcrelay}.
For the relay to remain live, every single block header between the latest state and target state must be submitted.
BTC relay incorporates an incentive mechanism by attaching fees to derived inclusion proofs and distributing them to block header relayers.
However, the mechanism requires high utilization, as many unused intermediary headers would have to be submitted in order to maintain live otherwise.
As a result, BTC relay has stalled for more than 38 months since the last block has been submitted at the time of writing. 

zkRelay is a chain relay that also aims for the submission of Bitcoin block headers to blockchains supporting the \gls{evm}, but significantly decreases execution costs compared to BTC relay by conducting off-chain header validation~\cite{westerkamp2020zkRelay}.
Block headers are combined in batches and verified in an off-chain program.
Correct program execution is ensured by utilizing zkSNARKs, so that the on-chain smart contract only verifies the correct program execution, rather than the header chain itself.
As the validation costs of the zkSNARK proof are independent of the off-chain program's complexity~\cite{Eberhardt2018}, its efficiency increases with large batch sizes.
While zkRelay provides significant improvements compared to BTC relay, it targets the validation of \gls{pow} and does not support \gls{pos} validation.

ETH relay constitutes a chain relay that links Ethereum with any blockchain supporting the \gls{evm}~\cite{frauenthaler2020ethrelay}.
Ethereum applies a memory-hard \gls{pow} algorithm to prevent centralization tendencies that originate from the use of customized hardware\cite{Wood2014}.
As on-chain validation of memory-hard algorithms is complex, ETH relay follows an optimistic approach and accepts block headers without verification.
Submitted headers are blocked for a contesting period in which it can be challenged by observers.
Only in case of a challenge, the submitted block header is validated on-chain and the submitter is slashed in case fraudulent behaviour is detected.
The optimistic approach taken by ETH relay potentially decreases operational costs significantly, as costly validation operations are only conducted in case of misbehaviour.
Therefore, while targeting \gls{pow} consensus algorithms in its current proposal, its general concept may be applicable to \gls{pos} chain relays as well.
Yet, as the concept depends on observing intermediaries and sufficient incentivization for correct behaviour, trust in transferred block headers is limited compared to full validation.

Ga\v{z}i et al. have proposed a sidechain mechanism for \gls{pos}-based blockchains that does validate block headers of the source blockchain, but operates on explicitly signed cross-chain transactions~\cite{Gazi2019}.
A subset of the source blockchain's validator is sampled to create certificates that notarize cross-chain transactions.
The solution is tailored to cater for Nakamoto-inspired \gls{pos} blockchains such as Cardano.
Thus, different assumptions are applied concerning finality and verifiability, compared to Verilay.
Furthermore, while generalizable, the concept enables a specific use-case of transferring tokens.
In contrast, Verilay applies the source blockchain's consensus to maintain the source blockchain's security properties and enables inclusion proofs of arbitrary data.





\section{Discussion}
The evaluation of Verilay demonstrates the concept's applicability under the assumption that signatures applied by the source blockchain's consensus protocol can be efficiently validated.
In case of Ethereum 1.0's \gls{evm}, this will be the case as soon as the respective \glspl{eip} are applied.
Furthermore, the number of signatures that must be validated needs to be limited.
Therefore, Ethereum 2.0's validator set cannot be efficiently validated and we rely on its light client protocol that provides a sub-sample of predefined size.
Since misbehavior of the sub-sample cannot be slashed, it is assumed that a majority of the sync committee acts altruistically honest.
Thus, the security guarantees are inferior compared to the validation of the full validator set.
Yet, some \gls{pos} protocols that do not offer accountable safety operate under similar assumptions~\cite{Badertscher2018OuroborosGenesis}.

Validator set changes constitute a challenge to Verilay if they are not efficiently verifiable.
As succeeding sync committees are signed by a trusted set in Ethereum 2.0, the chain relay need not re-execute the sampling.
\gls{pos} protocols that require all blocks of an epoch in order to calculate the next validator set may require the submission of all respective blocks to the chain relay.
Yet, this requirement may be alleviated if the source blockchain's protocol defines limited validator set changes.
\section{Conclusion}
In this paper, we proposed Verilay, a chain relay that provides interoperability between \gls{pbft}-inspired \acrlong{pos} blockchains and any blockchain that is capable of executing smart contracts.
We enable the submission of \gls{pos} block headers to a relay contract without requiring trust in the executing entity, as submitted block headers are verified based on the source blockchain's consensus protocol.
Since \gls{pbft}-based \gls{pos} protocols provide finality, no fork handling is required, and submitted blocks commit to the entire blockchain history.
Thus, Verilay requires only a single update during each validator set period, providing a lightweight solution compared to previous chain relay proposals.

The prototypical implementation of Verilay constitutes a chain relay for the Ethereum 2.0 beacon chain and provides a relay contract operating on the \gls{evm}.
Our evaluation shows that the proposed solution is viable without requiring changes to the consensus protocol of Ethereum 2.0 or the \gls{evm}.
Recent blocks can be submitted upon their finalization by the source blockchain's consensus.
Since only a single block must be submitted during each sync committee period, which lasts 17.3 hours, execution costs decrease significantly.
Hereby, Verilay demonstrates that \gls{pos} chain relays constitute a promising solution for blockchain interoperability.
\bibliographystyle{ieeetr}
\bibliography{references}
\vspace{12pt}

\end{document}

%% file: glossaries.tex
\newacronym{pow}{PoW}{Proof-of-Work}
\newacronym{pos}{PoS}{Proof-of-Stake}
\newacronym{dpos}{dPoS}{delegated Proof-of-Stake}
\newacronym{spv}{SPV}{Simplified Payment Verification}
\newacronym{asic}{ASIC}{Application-Specific Integrated Circuit}
\newacronym{nipopow}{NIPoPoW}{Non-Interactive Proofs of Proof-of-Work}
\newacronym{evm}{EVM}{Ethereum Virtual Machine}
\newacronym{pbft}{PBFT}{Practical Byzantine Fault Tolerance}
\newacronym{ssz}{SSZ}{SimpleSerialize}
\newacronym{bls}{BLS}{Boneh–Lynn–Shacham}
\newacronym{eip}{EIP}{Ethereum Improvement Proposal}

%% file: main.bbl
\begin{thebibliography}{10}

\bibitem{buterin2016interoperability}
V.~Buterin, ``{Chain Interoperability},'' 2016.
\newblock
  \url{https://www.r3.com/wp-content/uploads/2017/06/chain\_interoperability\_r3.pdf}
  Accessed: 2021-09-02.

\bibitem{heiss2019oracles}
J.~Heiss, J.~Eberhardt, and S.~Tai, ``{From Oracles to Trustworthy Data
  On-Chaining Systems},'' in {\em 2019 IEEE International Conference on
  Blockchain (Blockchain)}, pp.~496--503, 2019.

\bibitem{Meyden2019AtomicSwaps}
R.~v.~d. Meyden, ``On the specification and verification of atomic swap smart
  contracts (extended abstract),'' in {\em 2019 IEEE International Conference
  on Blockchain and Cryptocurrency (ICBC)}, pp.~176--179, 2019.

\bibitem{Xu2021AtomicSwaps}
J.~Xu, D.~Ackerer, and A.~Dubovitskaya, ``{A Game-Theoretic Analysis of
  Cross-Chain Atomic Swaps with HTLCs},'' in {\em 2021 IEEE 41st International
  Conference on Distributed Computing Systems (ICDCS)}, pp.~584--594, 2021.

\bibitem{westerkamp2020zkRelay}
M.~{Westerkamp} and J.~{Eberhardt}, ``{zkRelay: Facilitating Sidechains using
  zkSNARK-based Chain-Relays},'' in {\em 2020 IEEE European Symposium on
  Security and Privacy Workshops (EuroS\&PW)}, pp.~378--386, 2020.

\bibitem{btcrelay}
``{BTC Relay}.'' \url{https://github.com/ethereum/btcrelay}.
\newblock {Accessed: 2021-07-14}.

\bibitem{frauenthaler2020ethrelay}
P.~Frauenthaler, M.~Sigwart, C.~Spanring, M.~Sober, and S.~Schulte, ``{ETH
  Relay: A Cost-efficient Relay for Ethereum-based Blockchains},'' in {\em 2020
  IEEE International Conference on Blockchain (Blockchain)}, pp.~204--213,
  2020.

\bibitem{Gazi2019}
P.~Ga\v{z}i, A.~Kiayias, and D.~Zindros, ``{Proof-of-Stake Sidechains},'' in
  {\em 2019 2019 IEEE Symposium on Security and Privacy (SP)}, (Los Alamitos,
  CA, USA), pp.~677--694, IEEE Computer Society, may 2019.

\bibitem{Buterin2020POS}
V.~Buterin, D.~Hernandez, T.~Kamphefner, K.~Pham, Z.~Qiao, D.~Ryan, J.~Sin,
  Y.~Wang, and Y.~X. Zhang, ``{Combining {GHOST} and Casper},'' {\em CoRR},
  vol.~abs/2003.03052, 2020.

\bibitem{Wood2016}
G.~Wood, ``{Polkadot: Vision for a Heterogeneous Multi-chain Framework},''
  2016.
\newblock \url{https://polkadot.network/PolkaDotPaper.pdf} Accessed:
  2021-03-18.

\bibitem{buchman2019gossip}
E.~Buchman, J.~Kwon, and Z.~Milosevic, ``The latest gossip on {BFT}
  consensus,'' {\em CoRR}, vol.~abs/1807.04938, 2018.

\bibitem{nakamoto2008}
S.~Nakamoto, ``{Bitcoin: A Peer-to-Peer Electronic Cash System},'' 2008.
\newblock \url{http://www.bitcoin.org/bitcoin.pdf}. Accessed: 2021-05-02.

\bibitem{Badertscher2018OuroborosGenesis}
C.~Badertscher, P.~Ga\v{z}i, A.~Kiayias, A.~Russell, and V.~Zikas, ``Ouroboros
  genesis: Composable proof-of-stake blockchains with dynamic availability,''
  in {\em Proceedings of the 2018 ACM SIGSAC Conference on Computer and
  Communications Security}, CCS '18, (New York, NY, USA), p.~913–930,
  Association for Computing Machinery, 2018.

\bibitem{Buterin2017}
V.~Buterin and V.~Griffith, ``{Casper the Friendly Finality Gadget},'' {\em
  CoRR}, vol.~abs/1710.09437, 2017.

\bibitem{Kwon2014}
J.~Kwon, ``{Tendermint: Consensus without Mining},'' 2014.
\newblock
  \url{https://cdn.relayto.com/media/files/LPgoWO18TCeMIggJVakt\_tendermint.pdf}
  Accessed: 27.11.2018.

\bibitem{Kiayias2017Ouroboros}
A.~Kiayias, A.~Russell, B.~David, and R.~Oliynykov, ``Ouroboros: A provably
  secure proof-of-stake blockchain protocol,'' in {\em Advances in Cryptology
  -- CRYPTO 2017} (J.~Katz and H.~Shacham, eds.), (Cham), pp.~357--388,
  Springer International Publishing, 2017.

\bibitem{Deirmentzoglou2019Attack}
E.~Deirmentzoglou, G.~Papakyriakopoulos, and C.~Patsakis, ``{A Survey on
  Long-Range Attacks for Proof of Stake Protocols},'' {\em IEEE Access},
  vol.~7, pp.~28712--28725, 2019.

\bibitem{Stewart2020Grandpa}
A.~Stewart and E.~Kokoris{-}Kogia, ``{GRANDPA: a Byzantine Finality Gadget},''
  {\em CoRR}, vol.~abs/2007.01560, 2020.

\bibitem{Castro1999PBFT}
M.~Castro and B.~Liskov, ``Practical byzantine fault tolerance,'' in {\em
  Proceedings of the Third Symposium on Operating Systems Design and
  Implementation}, OSDI '99, (USA), p.~173–186, USENIX Association, 1999.

\bibitem{Braithwaite2020TendermintLightClient}
S.~Braithwaite, E.~Buchman, I.~Khoffi, I.~Konnov, Z.~Milosevic, R.~Ruetschi,
  and J.~Widder, ``A tendermint light client,'' {\em CoRR},
  vol.~abs/2010.07031, 2020.

\bibitem{Zamyatin2019}
A.~Zamyatin, N.~Stifter, A.~Judmayer, P.~Schindler, E.~Weippl, and W.~J.
  Knottenbelt, ``{A Wild Velvet Fork Appears! Inclusive Blockchain Protocol
  Changes in Practice},'' in {\em Financial Cryptography and Data Security},
  pp.~31--42, Springer Berlin Heidelberg, 2019.

\bibitem{Wood2014}
G.~Wood, ``{Ethereum: a secure decentralised generalised transaction ledger},''
  {\em Ethereum Project Yellow Paper}, 2021.
\newblock \url{https://ethereum.github.io/yellowpaper/paper.pdf}. Accessed:
  2021-07-05.

\bibitem{Ethereum2Specs2021}
``{Ethereum Proof-of-Stake Consensus Specifications},'' 2021.
\newblock \url{https://github.com/ethereum/consensus-specs}. Accessed:
  2021-11-17.

\bibitem{Cosmos2019Whitepaper}
J.~Kwon and E.~Buchman, ``{Cosmos - A Network of Distributed Ledgers},'' 2019.
\newblock \url{https://github.com/cosmos/cosmos/blob/master/WHITEPAPER.md}.
  Accessed: 2021-12-02.

\bibitem{Stelios2020LightClient}
S.~Daveas, K.~Karantias, A.~Kiayias, and D.~Zindros, ``A gas-efficient
  superlight bitcoin client in solidity,'' in {\em Proceedings of the 2nd ACM
  Conference on Advances in Financial Technologies}, AFT '20, (New York, NY,
  USA), p.~132–144, Association for Computing Machinery, 2020.

\bibitem{Boneh2001BLS}
D.~Boneh, B.~Lynn, and H.~Shacham, ``Short signatures from the weil pairing,''
  in {\em Advances in Cryptology --- ASIACRYPT 2001} (C.~Boyd, ed.), (Berlin,
  Heidelberg), pp.~514--532, Springer Berlin Heidelberg, 2001.

\bibitem{Eberhardt2018}
J.~{Eberhardt} and S.~{Tai}, ``{ZoKrates - Scalable Privacy-Preserving
  Off-Chain Computations},'' in {\em 2018 IEEE International Conference on
  Blockchain}, pp.~1084--1091, July 2018.

\end{thebibliography}
